\documentstyle[12pt]{article}
\begin{document}
\def\slash#1{#1 \hskip -0.5em / }
\global\arraycolsep=2pt 
\input{epsf}
\thispagestyle{empty}
\begin{titlepage}

\begin{flushright}
CERN-TH.7264/94   \\
hep-ph/9405353 
\end{flushright}

\vspace{0.3cm}

\begin{center}
{\Large\bf Weak Decays of Heavy--Quark Systems}
\end{center}

\vspace{0.8cm}

\begin{center}
Thomas Mannel  \\
{\sl Theory Division, CERN, CH-1211 Geneva 23, Switzerland}
\end{center}

\vspace{2.0cm}

\begin{abstract}
\noindent
The recent theoretical progress in the description of semileptonic 
decays in the framework of the heavy mass expansion is summarized. 
Both inclusive and exclusive decays are considered.
\end{abstract}
\vfill
\begin{center}
Contribution to the XXIXth Rencontres de Moriond, \\
            {\it Electroweak Interactions and Unified Theories}, \\ 
             M\'eribel, France, March 12--19, 1994.
\end{center}
\bigskip\bigskip
\vfill
\noindent
CERN-TH.7264/94\\
May 1994
\end{titlepage}
\setcounter{page}{1}
\section{Introduction}
Hadrons containing heavy quarks  
have attracted a lot of theoretical attention recently, since there has 
been an improvment in the  
understanding of heavy hadrons. The Standard
Model describes the decays of the heavy quarks, but due to the 
confinement property of the strong interactions only the decays of 
heavy hadrons  may be observed. In order to make contact between the 
quark and the hadron level, one has to deal with the 
QCD bound-state problem, which has not yet been solved. 
For this reason one has to use models to describe the weak 
decays of hadrons.  
 
However, it has been observed recently by various authors
\cite{HQL,HQET} that the 
model dependence may be reduced substantially    
for $b$ and $c$ quarks. The mass of these quarks is 
much larger than the energy scale determined by the 
light degrees of freedom, and it 
is convenient to switch to an effective theory description in which the 
dynamical degrees of freedom of the heavy quark are ``integrated out''; 
in other words, the heavy quark becomes an infinitely heavy, static
source of a colour field, moving with a fixed velocity.
 
This ``Heavy Quark Effective Theory'' (HQET) corresponds to a systematic
expansion of the full QCD Green-functions in inverse powers of the 
heavy-quark mass(es). The leading terms of this heavy mass expansion may 
be constructed from a renormalizable Lagrangian, which may be obtained 
from QCD by 
following the usual steps to construct an effective theory \cite{derivation}. 

In the heavy mass limit, two new symmetries appear, 
which are not present in QCD. The first symmetry is the heavy flavour 
symmetry, which is due to the fact that the interaction of the quarks with 
the gluons is flavour blind and in the heavy mass limit all heavy quarks 
act as a static source of colour. Formally this corresponds to an $SU(2)$
symmetry relating $b$ into $c$ quarks moving with the same velocity.  
The second symmetry is the spin symmetry of the heavy quark. 
The interaction of the heavy quark spin with 
the ``chromomagnetic'' field is inversely proportional to the heavy mass
and hence vanishes in the infinite mass limit. As a consequence, the 
rotations for the heavy quark spin become an $SU(2)$ symmetry, which 
holds for a fixed velocity of the heavy quark. 

Corrections to the limit $1/m_Q = 0$ may be studied systematically
in the framework of HQET. The corrections
are given as power series expansions in two small parameters. The
first small parameter is the strong coupling constant
taken at the scale of the heavy quark $\alpha_s (m_Q)$. This type
of correction may be calculated systematically using perturbation
theory in HQET. 
The second type of correction is characterized by the small parameter
$\bar\Lambda / m_Q$, where $\bar\Lambda$ is a scale of 
the light QCD degrees of freedom, e.g. $\bar\Lambda \sim m_{hadron}-m_Q$.
In the effective theory approach this type of correction enter 
through operators of higher dimension,
the matrix elements of which have to be parametrized in general
by additional form factors.

Inclusive decays may be treated in the heavy mass expansion by 
performing an operator product expansion much as is done in 
deep inelastic scattering \cite{isemi}. The leading term 
in this expansion 
turns out to be the decay of a free quark, and corrections may be 
parametrized by forward matrix elements  of higher dimensional 
operators. 

In this contribution I shall focus on semileptonic decays 
of heavy mesons and consider in the next section exclusive 
heavy-to-heavy transitions, with some emphasis on the model-independent 
determination of $V_{cb}$. In section 3 we summarize the recent 
developments in the application of the heavy mass expansion to 
inclusive semileptonic rates and decay spectra.

\section{Exclusive Semileptonic Decays}
The heavy-quark spin flavour symmetry 
strongly reduces the number of 
form factors in exclusive heavy-to-heavy transitions \cite{HQL}; in fact, 
for mesonic heavy-to-heavy transitions there is only one form factor, 
called the Isgur--Wise function. 
For a transition between
heavy ground state mesons ${\cal H}$ (either pseudoscalar or vector)
with heavy flavour $f$ ($f'$) moving with velocities $v$ ($v'$), one
obtains in the heavy-quark limit
\begin{equation} \label{WET}
\langle{\cal H}^{(f')} (v') | \bar{h}^{(f')}_{v'} \Gamma h^{(f)}_v
| {\cal H}^{(f)} (v) \rangle
= \xi (vv') C_\Gamma (v,v') .
\end{equation}
Here $h^{(f)}_v$ is the field operator annihilating a heavy quark of
flavour $f$, moving with velocity $v$; $\Gamma$ is some arbitrary Dirac
matrix. The Isgur--Wise function $\xi (vv')$ contains all the 
non-perturbative information for the 
heavy-to-heavy transition, while the coefficient  
$ C_\Gamma (v,v') $ may be calculated from the symmetries. 
Furthermore, heavy-quark symmetry fixes the value of $\xi$ at the
point $v = v'$ to be
\begin{equation} \label{norm}
\xi (vv' = 1) = 1 ,
\end{equation}
since the current $\bar{h}^{(f')}_{v} \Gamma h^{(f)}_v$ is one of the
generators of heavy-flavour symmetry.

In the following we shall treat the $b$ and the $c$ quarks as heavy. Then
(\ref{WET}) implies that the decays $B \to D \ell \bar{\nu}_\ell$ and 
$B \to D^* \ell \bar{\nu}_\ell$ are described by a single form factor, 
the Isgur--Wise function. In general, the relevant matrix elements are given
in terms of six form factors
\begin{eqnarray}
\langle D (v') | \bar{c} \gamma_\mu b | B(v) \rangle &=& \sqrt{m_B m_D}
\left[ \xi_+ (y) (v_\mu + v'_\mu)
     + \xi_- (y) (v_\mu - v'_\mu) \right] \\ \nonumber
\langle D^* (v',\epsilon) | \bar{c} \gamma_\mu b | B(v) \rangle &=&
       i \sqrt{m_B m_{D^*}}
\xi_V (y) \varepsilon_{\mu \alpha \beta \rho} \epsilon^{*\alpha}
                       v^{\prime \beta} v^\rho  \\ \nonumber
\langle D^* (v',\epsilon) | \bar{c} \gamma_\mu \gamma_5 b | B(v) \rangle &=&
       i \sqrt{m_B m_{D^*}}
\left[ \xi_{A1} (y) (vv'+1) \epsilon^*_\mu
      -  \xi_{A2} (y) (\epsilon^* v)  v_\mu \right. \\ \nonumber
&& \qquad \qquad \left.
      -  \xi_{A2} (y) (\epsilon^* v)  v'_\mu \right] ,
\end{eqnarray}
where we have defined $y = vv'$. In the heavy-quark limit, these
form factors are related to the Isgur--Wise function $\xi$ by
\begin{equation}
\xi_i (y) = \xi (y) \mbox{ for } i = +,V,A1,A3 , \quad
\xi_i (y) = 0       \mbox{ for } i = -,A2 .
\end{equation} 
 
The normalization statement (\ref{norm}) may be used to perform a
model-indepen\-dent determination of $V_{cb}$ from semileptonic 
heavy-to-heavy decays by extrapolating 
the lepton spectrum to the kinematic endpoint $v=v'$. 
Using the mode $B \to D^{(*)} \ell \nu$ one obtains the relation
\begin{equation} \label{extra}
\lim_{v \to v'} \frac{1}{\sqrt{(vv')^2-1}} \frac{d \Gamma}{d(vv')} =
\frac{G_F^2}{4 \pi^3} |V_{cb}|^2 (m_B - m_{D^*})^2 m_{D^*}^3 
|\xi_{A1} (1)|^2 .
\end{equation} 
In the heavy-quark limit the form factor $\xi_{A1}$ reduces to the 
Isgur--Wise function and is unity at the non-recoil point; aside from 
$|V_{cb}|$ everything in the right-hand side is known.

However, there are corrections to the normalization of the form 
factor $\xi_{A1}$ at zero recoil, which may be addressed using HQET. 
A complete discussion of the corrections
may be found in the review article by Neubert \cite{HQET}, including  
reference to the original papers. Here we only state the final 
result
\begin{eqnarray}
\xi_{A1} (1) &=& x^{6/25} \label{LLA} 
\left[ 1 + 1.561 \frac{\alpha_s (m_c) - \alpha_s (m_b)}{\pi}
              - \frac{8 \alpha_s (m_c)}{3 \pi} \right.  
\\
&& \left. + z \, \left\{\frac{25}{54} - \frac{14}{27} x^{-9/25} 
              + \frac{1}{18} x^{-12/25} + \frac{8}{25} \ln x 
       \right\}  
   - \frac{\alpha_s (\bar{m})}{\pi} \frac{z^2}{1-z} \ln z \right] 
\label{oneloop} \\
&& + \, \delta_{m^2}  
\label{msquared}
\end{eqnarray}
where we use the abbreviations
$$
x = \frac{\alpha_s (m_c)}{\alpha_s (m_b)} , 
\qquad
z = \frac{m_c}{m_b} , \qquad m_c < \bar{m} < m_b  .
$$

Contributions (\ref{LLA}) and (\ref{oneloop}) originate from QCD radiative 
corrections. Since for scales above $m_b$ and below $m_c$ the currents
are conserved, logarithmic corrections from running will only be induced
from scales between $m_b$ and $m_c$. The contribution (\ref{LLA}) 
is the correction in leading   
and next-to-leading logarithmic approximation;  
including the non-logarithmic one-loop contributions. 

Corrections of order $z = m_c / m_b$ may only be induced by QCD 
radiative corrections. The first term  in (\ref{oneloop}) is obtained 
by keeping the contributions of order $1/m_b$ in the matching at 
the scale $m_b$, performing the renormalization group running for these 
operators and matching at the scale $m_c$. In this way a resummation 
of logarithmic terms of the form $\alpha_s z \ln z$ is achieved. 
However, since $z \sim 0.3$ is not particularly small, 
one may as well choose to perform the matching from full
QCD to the effective theory in only one step, neglecting the running 
between $m_b$ and $m_c$. In this way one may retain the full dependence 
on $z$, at the price, however,of a scale ambiguity 
($m_c < \bar{m} < m_b$) in the choice of $\alpha_s$. The second correction
in (\ref{oneloop}) is obtained by this procedure, subtracting 
the linear term in $z$, which is already contained in the first term.

Finally there are recoil corrections to the normalization of $\xi_{A1}$.  
It has been shown \cite{Lu90} that the terms linear in $\Lambda / m_c$
and $\Lambda / m_c$ have to vanish due to heavy-quark symmetry. Thus 
the first non-vanishing recoil corrections are of order 
$(\Lambda / m_c)^2$, $(\Lambda / m_b)^2$ and $\Lambda^2 / (m_b m_c)$.
These contributions may only be estimated, since they need an input 
beyond heavy-quark effective theory. There are various estimates 
for these corrections, which are compatible with one another
$$ 
\delta_{m^2} = -2\% \ldots -3\% \,\, \mbox{\cite{FN92}}
, \quad
\delta_{m^2} =  0   \ldots -5\% \,\, \mbox{\cite{Ma94}}
, \quad
\delta_{m^2} =  0\% \ldots -8\% \,\, \mbox{\cite{SU94}}
$$
Adding all the corrections to the normalization, one obtains for the 
form factor $\xi_{A1}$
\begin{equation}
\xi_{A1} (1) = 0.96 \pm 0.03 ,
\end{equation}
where the error quoted is due to the uncertainty of the $1/m^2$ 
corrections. 

\begin{figure}[t]
   \vspace{0.5cm}
   \epsfysize=9cm
   \centerline{\epsffile{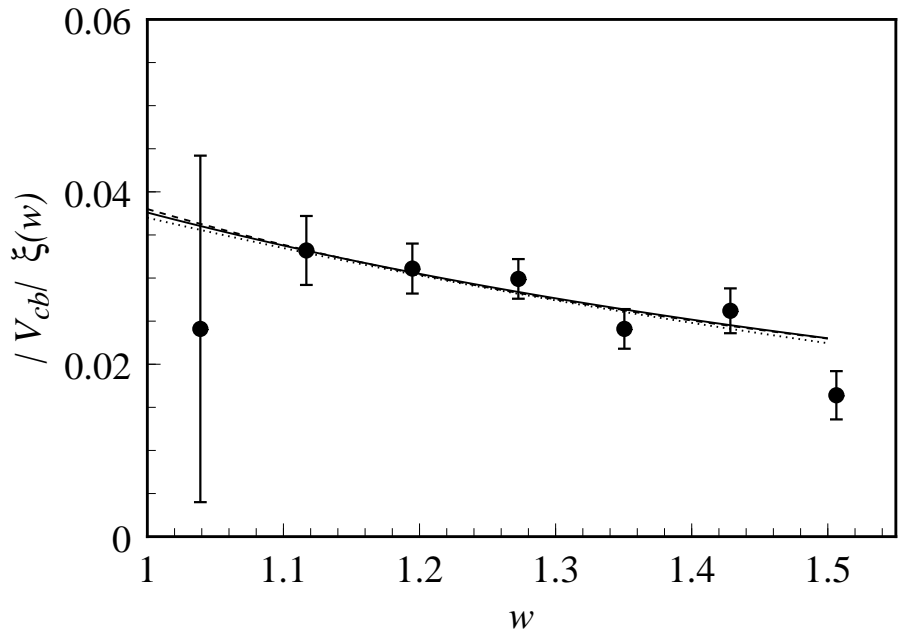}}
   \centerline{\parbox{11cm}{\caption{\label{fig1}
Latest data \protect{\cite{CLEO}} for the product $|V_{cb}| \xi (vv')$
as a function of $\omega = vv'$. Figure taken from \protect{\cite{NTASSI}}.
   }}}
\end{figure}

This result has been used to extract $V_{cb}$ from data. In Fig.~\ref{fig1}
the latest data \cite{CLEO} are shown. Three different forms 
of the Isgur--Wise function have been used for the extrapolation; however, the 
difference between the curves is almost invisible. From this fit one 
obtains 
\begin{equation}
|V_{cb}| \left(\frac{\tau_B}{1.5 \mbox{ps}}\right)^{1/2} = 
0.039 \pm 0.006 .
\end{equation}
This value of $|V_{cb}|$ is compatible with the value obtained in other
ways, e.g. from inclusive decays \cite{stone}.
 
\section{Inclusive Decays of Heavy Mesons}
In order to obtain a $1/m_Q$ expansion for inclusive decay rates one
uses an operator product expansion \cite{isemi}
as in deep inelastic scattering. Starting from the 
effective Hamiltonian
\begin{equation}
H_{eff} = \frac{G_F}{\sqrt{2}} V_{bq} (\bar b \gamma_\mu (1-\gamma_5) q)  
                                      (\bar \ell \gamma^\mu (1-\gamma_5) \nu) ,
\end{equation}
one writes the total inclusive rate as
\begin{equation} \label{incsetup}
\Gamma  =  \frac{1}{2 m_B} \sum_X (2 \pi)^4 \delta^4 (P_B - q - P_X )
| \langle X_s | H_{eff} | B(v) \rangle |^2 . 
\end{equation}
The matrix element appearing in (\ref{incsetup}) contains a large 
scale, the mass of the $b$ quark. To make the dependence on this scale
explicit, one redefines the $b$ quark field by removing a phase 
factor corresponding to an on-shell $b$ quark moving with the velocity
of the meson
\begin{equation} \label{preredef}
b(x) = \exp (-im_b vx) Q_v (x) .
\end{equation}
Inserting this in (\ref{incsetup}),  
\begin{equation}  
\Gamma =  \frac{1}{2 m_B}
\int d^4 x  \exp (im_b vx)\langle B(v) | \widetilde{H}_{eff} (x)
             \widetilde{H}_{eff}^\dagger (0) | B(v) \rangle  , 
\end{equation}
where the tilde denotes the effective Hamiltonian with $b$ replaced
by $Q_v$. Once the phase factor is extracted from the matrix element,
it no longer depends on the large mass, and a 
short-distance expansion may be performed for the operator product 
appearing in the 
matrix element. The relevant momentum is $m_b v$ and the short-distance
expansion has the form
$$
\int d^4 x  e^{im_b vx}
T \left[ \widetilde{H}_{eff} (x) \widetilde{H}_{eff}^\dagger (0) \right]
= \sum_{n=0}^\infty  \left(\frac{1}{2m_b}\right)^n 
     C_{n+3} (\mu) {\cal O}_{n+3} (\mu) ,
$$
where ${\cal O}_n$ are operators of dimension $n$, renormalized at scale 
$\mu$, and $C_n$ are the 
corresponding Wilson coefficients. 

The lowest-order term of the operator product expansion is the  
dimension-three operator ${\cal O}_3 = \bar Q_v Q_v$, and its forward
matrix element is normalized because of heavy-quark symmetries. 
Evaluating this contribution yields the free-quark decay rate. 

All dimension-four operators are proportional to the 
equations of motion ${\cal O}_4 \propto \bar Q_v (iv D) Q_v$, 
and the first non-trivial contribution comes from
dimension-five operators and are of order of $1/m_b^2$. 
For mesonic decays there are two matrix elements of dimension-five
operators:
$$
\langle B(v) | \bar{h}_v^{(b)} (i D)^2 h_v^{(b)} | B(v) \rangle 
= 2 m_b \lambda_1
\quad \mbox{and} \quad
\langle B(v) | \bar{h}_v^{(b)} i g \sigma_{\mu \nu} G^{\mu \nu} h_v^{(b)}
| B(v) \rangle = 12 m_b \lambda_2 ,
$$
where $4 \lambda_2 = m_{B^*}^2 - m_B^2$. The parameter $\lambda_1$
is not as easily accessible, but the expectations from  QCD
sum rules are $  \lambda_1  = - 0.52 \pm 0.12 $~GeV${}^2$ \cite{BB93}. 
 
In terms of these two parameters the non-perturbative corrections 
to the inclusive decay $B \to X_u \ell \nu$ is given by the expression
\begin{equation}
\Gamma(B \to X_u \ell \nu) = \Gamma_b 
 \left[
      1+\frac{\lambda_1 - 9 \lambda_2}{2m_b^2} 
      \right] ,  \qquad 
      \Gamma_b = \frac{G_F^2 m_b^5}{192\pi^3} |V_{ub}|^2 ;
\end{equation}
a similar expression may be obtained for the rate for 
$B \to X_c \ell \nu$. 

The spectrum of the charged lepton in inclusive decays may be calculated 
by similar means. The rate is written as a product of the hadronic and
leptonic tensor
\begin{equation}
d \Gamma = \frac{G_F^2}{4 m_B} | V_{Qq} |^2 W_{\mu \nu}  
\Lambda^{\mu \nu} d(PS) , 
\end{equation}
where $d(PS)$ is the phase-space differential.  
The operator product 
expansion along the lines described above 
is then performed for the two currents appearing in 
the hadronic tensor. Redefining the heavy-quark fields as in 
(\ref{preredef}) one finds that the momentum transfer variable 
relevant for the short-distance expansion is $m_b v - q$, where 
$q$ is the momentum transfer to the leptons. 

Again the contribution of the dimension-three operators yields the 
free-quark decay spectrum, and there are no contributions from 
dimension-four
operators. The $1/m_b^2$ corrections are parametrized in terms of 
$\lambda_1$ and $\lambda_2$ and the result for the lepton spectrum 
in  $B \to X_u \ell \nu$ is given by
\begin{eqnarray} 
\frac{1}{\Gamma_b} \frac{d\Gamma}{dy} &=& 2y^2 (3-2y)
        + \frac{10y^2}{3} \frac{\lambda_1}{m_b^2}
        + 2y(6+5y) \frac{\lambda_2}{m_b^2}  \nonumber \\
 && - \frac{\lambda_1 + 33 \lambda_2}{3m_b^2} \delta (1-y)
  -  \frac{\lambda_1}{3m_b^2} \delta ' (1-y) , \label{npc}
\end{eqnarray}
where $y = 2 E_\ell / m_b $ is the rescaled energy of the charged lepton.  

Aside from regular terms, the result also exhibits $\delta$-function 
singularities at the endpoint, indicating that higher terms in the 
operator product expansion become important; here $(m_b v - q)^2$ 
becomes small. 

\begin{figure}[t]
   \vspace{0.5cm}
   \epsfysize=9cm
   \centerline{\epsffile{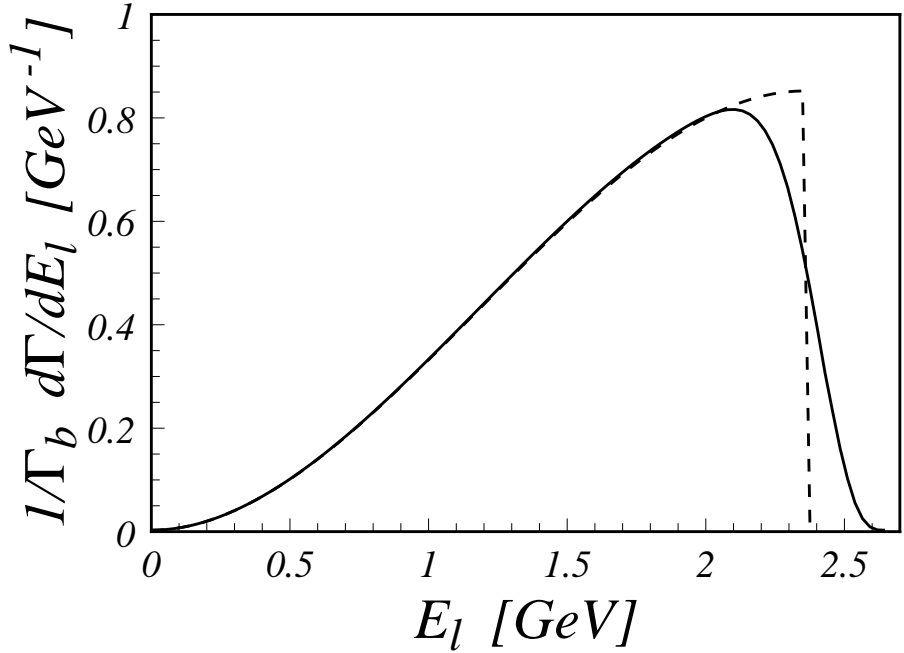}}
   \centerline{\parbox{11cm}{\caption{\label{fig2}
Charged-lepton spectrum in $B\to X_u \ell  \bar\nu$ decays. 
The solid line is (\protect{\ref{buconv}}) with the ansatz 
(\protect{\ref{ftoy}}), the dashed
line shows the prediction of the free-quark decay model.
   }}}
\end{figure}

It has been shown \cite{shape}
that the most singular terms of the short-distance expansion may be 
resummed into a structure function, analogous to a parton distribution 
function known from deep inelastic scattering. It is defined 
formally as 
\begin{equation}\label{fdef}
   f(k_+) = \frac{1}{2m_B} 
   \langle B(v)|\,\bar h_v\,\delta(k_+-i D_+)\,h_v\,
   |B(v)\rangle \quad \mbox{with} \quad k_+ = k_0 + k_3 .
\end{equation}
Using this function, the result for the spectrum becomes 
\begin{eqnarray} \label{buconv}
   \frac{d\Gamma}{d E_\ell} 
   &=& \frac{G_F^2 |V_{ub}|^2}{12\pi^3} E_\ell^2
       (3 m_b - 4 E_\ell) \int\limits_{2 E_\ell-m_b}^{\bar\Lambda}
       d k_+ \, f(k_+)  , 
\end{eqnarray}
where $\bar\Lambda = m_B - m_b$. The function $f$ is genuinely 
non-perturbative; in order to illustrate its effect, we choose a simple 
one-parameter model 
\begin{equation} \label{ftoy}
   f(k_+) = {32\over\pi^2\bar\Lambda}\,(1-x)^2\,
   \exp\bigg\{ - {4\over\pi}\,(1-x)^2 \bigg\}\,
   \Theta(1-x) \,;\quad x = {k_+\over\bar\Lambda} \,,
\end{equation}
and Fig.~\ref{fig2} shows the resulting spectrum for 
$\bar\Lambda = 570$ GeV.
Due to non-perturbative
effects, the spectrum now extends beyond the parton model endpoint $m_b / 2$
up to the ``physical'' endpoint $m_B / 2$. However, before these results may 
be confronted with data, the QCD radiative corrections need to be taken
into account. First results have been published 
\cite{rc}, but the issue is still in a state of flux.


\end{document}